\documentclass[12pt,tightenlines]{revtex4}
\usepackage{amsmath,amssymb}
\usepackage{psfrag}
\usepackage[dvips]{graphicx}

\begin{document}

\title{Reducing the mirrors coating noise in laser gravitational-wave
antennae by means of double mirrors}

\author{F.Ya.Khalili}

\email{farid@hbar.phys.msu.ru}

\affiliation{Moscow State University}


\begin{abstract}

Recent researches show that the fluctuations of the dielectric mirrors coating
thickness can introduce a substantial part of the future laser
gravitational-wave antennae total noise budget. These fluctuations are
especially large in the high-reflectivity end mirrors of the Fabry-Perot
cavities which are being used in the laser gravitational-wave antennae.

We show here that the influence of these fluctuations can be substantially
decreased by using additional short Fabry-Perot cavities, tuned in
anti-resonance instead of the end mirrors.

\end{abstract}

\maketitle

\section{Introduction}

One of the basis components of laser gravitational-wave antennae
\cite{Abramovici1992, Abramovici1996, WhitePaper1999} are high-reflectivity
mirrors with multilayer dielectric coating. Recent researches \cite{Levin1998,
Crooks2002, Harry2002, Nakagava2002, Penn2003, 03a1BrVy, 03a1BrSa,
Cagnoli2003, Fejer2004, Harry2004} have shown that fluctuations of the coating
thickness produced by, in particular, Brownian and thermoelastic noise in a
coating, can introduce substantial part of the total noise budget of the
future laser gravitational-wave antennae. For example, estimates, done in
\cite{03a1BrVy} show that the thermoelastic noise value can be close to the
Standard Quantum Limit (SQL) \cite{03a1BrGoKhMaThVy} which corresponds to the
sensitivity level of the Advanced LIGO project \cite{WhitePaper1999} or even
can exceed it in some frequency range.

For this reason it was proposed in \cite{04a1BrVy} to replace end mirrors by
coatingless corner reflectors. It was shown in this article that by using
these reflectors, it is possible, in principle, to obtain sensitivity much
better than the SQL. However, the corner reflectors require substantial
redesign of the gravitational-wave antennae core optics and suspension system.

At the same time, the value of the mirror surface fluctuations depends on the
number of dielectric layers which form the coating. It can be explained in the
following way. The most of the light is reflected from the first couple of the
layers. At the same time, fluctuations of the mirror surface are created by
the thickness fluctuations of all underlying layers, and the larger is the
layers number, the larger is the surface noise.

Therefore, the surface fluctuations are relatively small for the input mirrors
({\sf ITM}) of the Fabry-Perot cavities of the laser gravitational-wave
antennae with only a few coating layers and $1-{\cal R}\sim 10^{-2}$ (${\cal
R}$ is the mirror power reflectivity), and is considerably larger for the end
mirrors ({\sf ETM}) with coating layers number $\sim 40$ and $1-{\cal
R}\lesssim 10^{-5}$.

\begin{figure}

\psfrag{L}[ct][lb]{$L=4\,{\rm Km}$}
\psfrag{l}[ct][lb]{$l\lesssim 10\,{\rm m}$}
\psfrag{ITM}[cb][lb]{{\sf ITM}}
\psfrag{IETM}[cb][lb]{{\sf IETM}}
\psfrag{EETM}[cb][lb]{{\sf EETM}}

\begin{center}\includegraphics[width=5in]{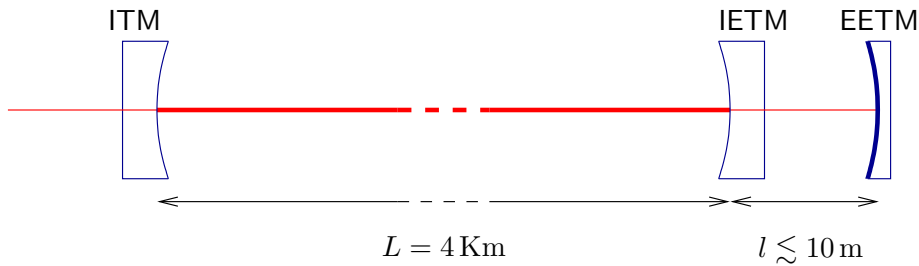}\end{center}

\caption{Schematic layout of a Fabry-Perot cavity with double mirror system
instead of the end mirror: {\sf ITM} and {\sf IETM} are similar moderate
reflective mirrors; {\sf EETM} is a high-reflective
one.}\label{fig:fabry_dbl_mirror}

\end{figure}

In this paper another, less radical way of reducing the coating noise,
exploiting this feature, is proposed. It is based on the use of an additional
short Fabry-Perot cavity instead of the end mirror (see
Fig.\,\ref{fig:fabry_dbl_mirror}). It should be tuned in anti-resonance, {\em
i.e} its optical length $l$ should be close to $l=(N+1/4)\lambda$, where
$\lambda$ is a wavelength. The back side of the first mirror have to have a
few layers of an antireflection coating.

It can be shown that in this case reflectivity of this cavity will be defined
by the following equation:

\begin{equation}\label{R_simple}
  1-{\cal R} \approx \frac{(1-{\cal R}_1)(1-{\cal R}_2)}{4} \,,
\end{equation}
where ${\cal R}_{1,2}$ are the reflectivities of the first ({\sf EETM} on
Fig.\,\ref{fig:fabry_dbl_mirror}) and the second ({\sf IETM}) mirrors. Phase shift
in the reflected beam produced by small variations $y$ in position of the
second mirror reflecting surface relative to the first one will be equal to

\begin{equation}\label{phi_simple}
  \phi \approx \frac{1-{\cal R}_1}{4}\times 2ky \,,
\end{equation}
where $k=2\pi/\lambda$ is a wave number. It is supposed for simplicity that
there is no absorption in the first mirror material; more general formulae are
presented below.

It follows from these formulae that the first mirror can have a moderate value
of reflectivity and, therefore, a small number of coating layers. In
particular, it can be identical to the input mirror of the main
Fabry-Perot cavity ({\sf ITM}). At the same time, influence of the coating noise
of the second (very-high-reflective) mirror will be suppressed by a factor of
$(1-{\cal R}_1)/4$, which can be as small as $\sim 10^{-2}\div 10^{-3}$.

\begin{figure}

\psfrag{ETM}[cb][lb]{{\sf ETM}}

\begin{center}\includegraphics[width=2in]{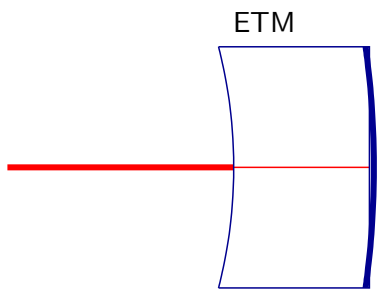}\end{center}

\caption{The double reflector based on a single
mirror.}\label{fig:single_mirror}

\end{figure}

In principle, another design of the double reflector is possible, which
consists of one mirror only, see Fig.\,\ref{fig:single_mirror}. Both surfaces
of this mirror have to have reflective coatings: the thin one on the face side
and the thick one on the back side. In this case the additional Fabry-Perot
cavity is created {\em inside} this mirror. However, in this case
thermoelastic fluctuations of the the back surface coating will bend the
mirror and thus will create unacceptable large mechanical fluctuations of the
face surface. Estimates show that using this design, it possible to reduce the
face surface fluctuations by factor $\sim 3$ only \cite{vyat_priv}. So the
design with two {\em mechanically isolated} reflectors only will be considered
here.

In the next section more detail analysis of this system is presented.

\section{Analysis of the double-mirror reflector}

\begin{figure}

\psfrag{a}[rc][lb]{$a$}
\psfrag{b}[rc][lb]{$b$}
\psfrag{a0}[lc][lb]{$a_0$}
\psfrag{b0}[lc][lb]{$b_0$}
\psfrag{a1}[lc][lb]{$a_1$}
\psfrag{b1}[lc][lb]{$b_1$}
\psfrag{a2}[rc][lb]{$a_2$}
\psfrag{b2}[rc][lb]{$b_2$}
\psfrag{IETM}[cb][lb]{{\sf IETM}}
\psfrag{EETM}[cb][lb]{{\sf EETM}}

\begin{center}\includegraphics[width=3.5in]{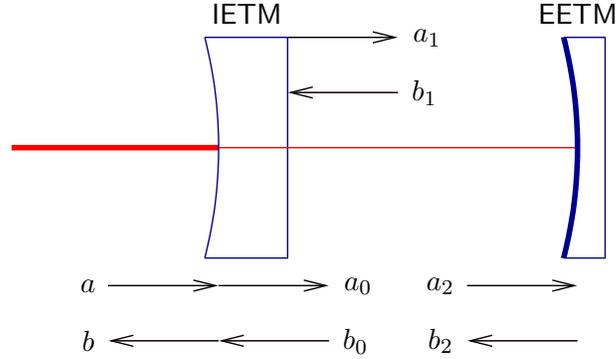}\end{center}

\caption{The double mirror reflector.}\label{fig:dbl_mirror}

\end{figure}

The rightmost part of Fig.\,\ref{fig:fabry_dbl_mirror} is presented in
Fig.\,\ref{fig:dbl_mirror}, where the following notation is used:

$a, b$ are the amplitudes of the incident and reflected waves for the first
mirror, respectively;

$a_0, b_0$ are the amplitudes of the waves traveling in the left and right
directions, respectively, just behind the first mirror coating;

$a_1, b_1$ are the same for the waves just behind the first mirror
itself;

$a_2, b_2$ are the amplitudes of the incident and reflected waves for the second mirror,
respectively.

These amplitudes satisfy the following equations:

{\allowdisplaybreaks\begin{subequations}\label{main_eqs}
  \begin{align}
    a_0 &= -R_1b_0 + iT_1a \,, \\
    a_1 &= T_0a_0 + A_1n_a \,, \\
    a_2 &= \theta a_1 \,, \\
    b &= -R_1a + iT_1b_0 \,, \\
    b_0 &= T_0b_1 + A_0n_b \,, \\
    b_1 &= \theta b_2 \,, \\
    b_2 &= -R_2a_2 + A_2n_2 \,,
  \end{align}
\end{subequations}}
where:

$n_a,n_b,n_2$ are independent zero-point oscillations generated in the
first ($n_a,n_b$) and the second ($n_2$) mirrors;

$\theta = e^{ikl_1}$, where $l_1$ is the distance between the first mirror
back surface and the second mirror;

$-R_1$ and $iT_1$ are the amplitude reflectivity and transmittance of the
first mirror coating, respectively, $R_1^2+T_1^2=1$;

$T_0$ and $A_0$ are the amplitude transmittance and absorption of the first
mirror bulk, respectively, $|T_0|^2+A_0^2=1$;

$-R_2$ and $A_2$ are the amplitude reflectivity and absorption of the second
mirror, respectively, $R_2^2+A_2^2=1$.

$R_1,T_1,A_0,R_2,A_2$ are real values; $T_0$ is a complex one, its argument
corresponds to the phase shift in the first mirror bulk.

Here we do not consider absorption in the first mirror coating for two
reasons: (i) it is relatively small and (ii) it exists both in traditional
one-mirror reflectors and in the one considered here, and the main goal of
this short article is to emphasize the {\em differences} between these two types
of reflectors.

We also suppose that the mirrors move rather slowly:

\begin{equation}
  \frac{dl}{dt} \ll \frac{l}{c} \,.
\end{equation}
In the case of the gravitational-wave signal characteristic frequencies
$\Omega\lesssim 10^3\,{\rm s}^{-1}$ and relatively short length $l\lesssim
1\,{\rm m}$ this inequality is fulfilled pretty well.

It follows from equations (\ref{main_eqs}) that the reflected beam
amplitude is equal to

\begin{equation}\label{dbl_soln_b}
  b = \frac{(R_2T_0^2\theta^2-R_1)a - iR_2A_0T_0T_1\theta^2n_a
    + iA_2T_0T_1\theta n_2 + iA_0T_1n_b}{1-R_1R_2T_0^2\theta^2} \,.
\end{equation}
This solution can be presented in the following form:

\begin{equation}
  b = \tilde R a + A n \,,
\end{equation}
where

\begin{equation}
  \tilde R = \frac{R_2T_0^2\theta^2 - R_1}{1 - R_1R_2T_0^2\theta^2}
\end{equation}
is the equivalent complex reflection factor for the scheme considered,

\begin{equation}
  A = \frac{T_1\sqrt{1 - R_2^2|T_0|^4}}{|1-R_1R_2T_0^2\theta^2|}
\end{equation}
is its equivalent absorption factor, and

\begin{equation}
  n = \frac{1}{A}\,\frac{ - iR_2A_0T_0T_1\theta^2n_a
    + iA_2T_0T_1\theta n_2 + iA_0T_1n_b}{1-R_1R_2T_0^2\theta^2}
\end{equation}
is the sum noise normalized as zero-point fluctuations.

As mentioned above, this system should be tuned in anti-resonance:

\begin{equation}
  l \equiv \frac{1}{k}\arg{T_0\theta}
  = \frac{\pi}{k}\left(N + \frac{1}{2}\right) + y \,, \\
\end{equation}
where $N$ is an integer and $y \ll \lambda$. In this case

\begin{equation}
  T_0\theta = i(-1)^N|T_0|e^{iky} \approx i(-1)^N(1+iky) \,,
\end{equation}
and

\begin{equation}
  \tilde R \approx -Re^{i\phi} \,,
\end{equation}
where

\begin{equation}
  R = 1-\frac{(1-R_1)(1-R_2|T_0|^2)}{1 + R_1R_2|T_0|^2}\,,
\end{equation}
and

\begin{equation}
  \phi \approx
    \frac{2R_2|T_0|^2T_1^2}{(R_2|T_0|^2 + R_1)(1 + R_1R_2|T_0|^2)}\,ky
\end{equation}
is the phase shift produced by the deviation $y$ in the distance $l$.

Suppose that factors $T_1, A_0, A_2$ are small. In this case

\begin{equation}
  1-R \approx \frac{(1-R_1)(1-R_2+A_0^2)}{2} \label{R_amp} \,, \\
\end{equation}
\begin{equation}
  \phi \approx (1-R_1)ky \,. \label{phi_amp}
\end{equation}
Using power reflection and absorption factors instead of the amplitude ones:

\begin{gather}
  {\cal R} = R^2 \,, \\
  {\cal R}_{1,2} = R_{1,2}^2 \,, \\
  {\cal A}_0 = A_0^2 \,,
\end{gather}
equations (\ref{R_amp}), (\ref{phi_amp}) can be rewritten as follows:

\begin{equation}
  1- {\cal R} \approx \frac{(1-{\cal R}_1)(1-{\cal R}_2 + 2{\cal A}_0)}{4} \,,
    \label{R_pow} \\
\end{equation}
\begin{equation}
  \phi \approx \frac{1-{\cal R}_1}{2}\,ky \,.
\end{equation}

\section{Conclusion}

The main goal of this short article is just to claim the idea, so the detailed
design of the additional cavity is not presented here. However, the following
important topics have to be discussed in brief.

The first one concerns the optimal value of the {\sf IETM} mirror
reflectivity. The smaller is $1-{\cal R}_1$, the larger is suppression factor
for the {\sf EETM} mirror surface noises; at the same time, the larger is the
{\sf IETM} mirror coating noise. The rigorous optimization requires exact
knowledge of the coating noise dependence on the coating layers number.

A crude estimate based in the exponential dependence of the {\sf IETM} mirror
transmittance ${\cal T}_1 \approx 1-{\cal R}_1$ on the coating layers number
gives that the optimal transmittance value is relatively large, ${\cal T}_1
\sim 10^{-1}$. On the other hand, smaller values of the {\sf IETM} mirror
transmittance, down to the input ({\sf ITM}) mirror transmittance ${\cal
T}_{\sf ITM}$ are also acceptable. Therefore, identical {\sf ITM} and {\sf
IETM} mirrors can be used.

In the Advanced LIGO interferometer, the input mirrors transmittance will be
equal to ${\cal T}_{\sf ITM}\approx 5\times 10^{-3}$, and its bulk absorption
will be equal to ${\cal A}_{\rm ITM} \approx 10^{-5}$ \cite{RefDesign}. Using
such mirror as an {\sf IETM} mirror in the scheme proposed in this article,
and mirror with commercially available value of $1-{\cal R}_2 \approx 10^{-5}$
as an {\sf EETM} mirror, it is possible to create a double-mirror reflector
with $1-{\cal R} < 10^{-6}$ and suppression factor for the {\sf EETM} surface
fluctuations $\dfrac{1-{\cal R}_1}{4} \approx 10^{-3}$.

The second issue concerns the optical power circulating through the {\sf IETM}
mirror. It is easy to show using equations (\ref{main_eqs}), that it is
$\dfrac{4}{1-{\cal R}_1}\sim 10^3$ times smaller that the power circulating in
the main cavities. In the Advanced LIGO topology, it will be approximately
equal to the power circulating through the {\sf ITM} mirrors and the
beamsplitter (about 1\,KW).

It is necessary to note also that $y$ in the calculations presented above
includes not only coating noise of the {\sf EETM} mirror but all possible
kinds of its surface fluctuations, including ones caused by Brownian and
thermoelastic fluctuations in this mirror bulk, Brownian fluctuation in its
suspension, seismic noise as well as the mirror quantum fluctuations. This
feature simplifies greatly the {\sf EETM} mirror design because the
requirements for all these noise sources can be reduced by a factor of
$(1-{\cal R}_1)/4$.

In particular, the SQL value $\sqrt{\dfrac{\hbar}{m\Omega^2}}$ for this mirror
($m$ is its mass and $\Omega$ is the observation frequency) can be larger by
a factor of $\left(\dfrac{1-{\cal R}_1}{4}\right)^{-1}$. Therefore, its mass
can be, in principle, $\left(\dfrac{1-{\cal R}_1}{4}\right)^{-2} \sim 10^6$
times smaller than for the main ({\sf ITM} and {\sf IETM}) mirrors.
Of course, such a small mirror hardly can be used in the real interferometer.
This estimates shows only that the quantum noise does not impose any practical
limitation on the ${\sf EETM}$ mirror mass.

\section*{Acknowledgments}

The author is grateful to V.B.Braginsky, S.L.Danilishin, G.Harry, D.Ottaway,
D.Shoemaker and S.P.Vyatchanin for useful remarks.

This work was supported by NSF grant PHY0098715, by Russian Ministry of
Industry and Sciences contracts 40.02.1.1.1.1137 and 40.700.12.0086, and by
Russian Foundation of Basic Researches Grant 03.02.16975-a.


\end{document}